# Spectral Response of Al/Si Photodiode as IR Sensor

**Irzaman, A. Fuad, D. Rusdiana, H. Saragih, T. Saragi, M. Barmawi**
*Department of Physics, FMIPA ITB, Jl. Ganesha 10 Bandung, Indonesia*
*E-mail :* irzaman@student.fi.itb.ac.id

July 11, 2001

## ABSTRACT

*Al/Si photodiode were prepared by successive of deposition of Aluminum thin films on p-type Si (100) using coprecipitation method by evaporator in vacuum $2 \times 10^{-4}$ torr. The measured signals are attributed to a spectral response, which appears in the studied photodiode. The photodiode Al/Si is sensitivity over a broad range of IR wavelengths from 800 to 1000 nm, when illuminated with discontinuous light at chopper frequency 200 Hz. The spectral response distribution that sensitivity peaks appear when the measurement was performed using the ac mode. The peak was at 981 nm, the incident radiation power ($P_i$) at the surface was 25 $\mu W/cm^2$, the current signal ($I_p$) was 12.2 nA, the quantum efficiency ($\eta$) was 6.16 % and the current sensitivity was 0.488 $nA/\mu W/cm^2$.*

## 1. Introduction

The infrared (IR) photodiode has advantages of waveligth-independent sensitivity and room temperature operation, and it is expected to provide various thermal observations for objects at near ambient temperature [1]. Thin films of metal-semiconductor photodiodes are particularly useful in the ultraviolet, visible and IR light regions. A Au/Si photodiode having 100 Å and 500 Å zinc sulfide as the antireflection coating, more then 95 % of the incident light with $\lambda$ = 6328 Å (red light) will be transmitted into the silicon substrate, and quantum efficiency is 80 % [2].

Pt/Si photodiode were prepared by successive of deposition of platinum thin films on Si (100) using reactive magnetron sputtering [3]. Characteristics of photodiode from RS Components catalogue to size: the peak spectral response is at 750 nm (IR wavelength), the sensitivity is quoted 0.7 $nA/\mu W/cm^2$, the typical dark current at bias of – 20 volt is 1.4 nA, which means that the minimum detectable power input us of the order of 2 $\mu W/cm^2$ [4].

## 2. Basic Theory

A photodiode is a type of photoresistor in which the incident light falls on a semiconductor junction, and the separation of electron and holes caused by the action of light will allow the junction to conduct when it is reverse-biased. Al/Si photodiode were prepared by successive of deposition of Aluminum thin films on *p*-type Si (100) using coprecipitation method by evaporator in vacuum $2 \times 10^{-4}$ torr. The construction of Al/Si photodiode is shown in Figure 1.

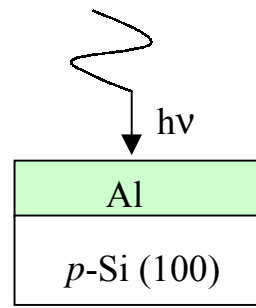

**Figure 1.** Construction of Al/Si photodiode.

The spectral distribution measurement were performed using monochromator light, within a broad range of wavelengths form 800 to 1000 nm, a SR530 Lock in Amplifier and chopper was reference (200 Hz). For the infrared range of the electromagnetic spectrum a Xe lamp model Spectral Energy LPS255HR Universal Arc. Lamp, a Spex 1681 monochromator, and Luxmeter Model LX 101 were used.

## 3. EXPERIMENTAL RESULT

An intense source of IR light of wavelength 800 to 1000 nm is provided by a Xe lamp model Spectral Energy LPS255HR Universal Arc. Lamp to pass a Spex 1681 monochromator. Each electrode is a filament for providing electrons to maintain an electric discharge through an inert gas. After the inert gas discharge has taken place for a few minutes, the temperature rises to a value at which the vapor pressure of the xenon is great enough to provide sufficient xenon atoms to emit the characteristic IR light.

In a Xe lamp electrical power is supplied and radiation is emitted. The radiant energy emitted per unit of time is called radiant power or radiant flux (watt). Not all the electrical power is converted into radiant power, some is lost by heat conduction, heat convection and absorption. Luminous efficiency expresses a property of a sample of radiant power.

Illumincance (lux) is luminous flux (lumen) incident per unit area ($m^2$) can be expressed as Eq. (1) :

$$E = \frac{\Delta F}{\Delta A}. \qquad (1)$$

Approximate conversion illuminance to incident power can be expressed as eq. (2) : [4]

$$200 \text{ lux} = 1 \text{ mW/cm}^2 \qquad (2)$$

Figure 1 shows the illuminance versus IR wavelength and the incident power versus IR wavelength from a Xe lamp at power source = 260 watt, at Al/Si photodiode surface (A) = 1 mm$^2$ using Luxmeter Model 101, resulted the average illuminance was 5 lux, means the average incident power was 25 µW/cm$^2$ with using Eq. (2).

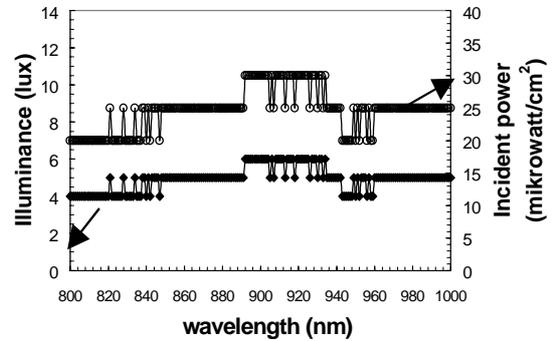

**Figure 2.** The illuminance versus IR wavelength and the incident power versus IR wavelength from from a Xe lamp at power source = 260 watt, at Al/Si photodiode surface (A) = 1 mm$^2$ using Luxmeter Model 101.

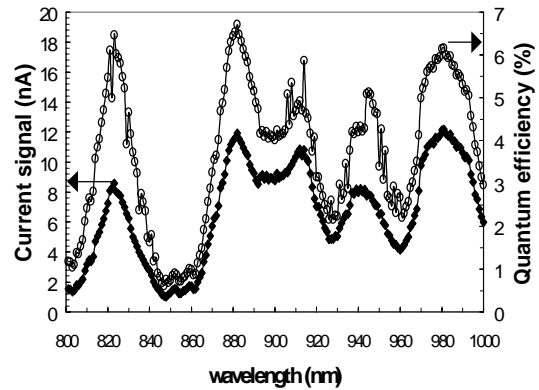

**Figure 3.** The current signal versus IR wavelength and the quantum efficiency versus IR wavelength at chopper frequency = 200 Hz.

The quantum efficiency ($\eta$) as mentioned previously is the number of electron – hole pairs generated for each incident photon expressed as Eq. (3) : [2,5]

$$\eta = \left(\frac{I_p}{q}\right)\left(\frac{h\nu}{P_i}\right) \qquad (3)$$

where $I_p$ is the photogenerated output current from the absorption of incident optical power $P_i$ at a wavelength (corresponding to photon energy $h\nu$).

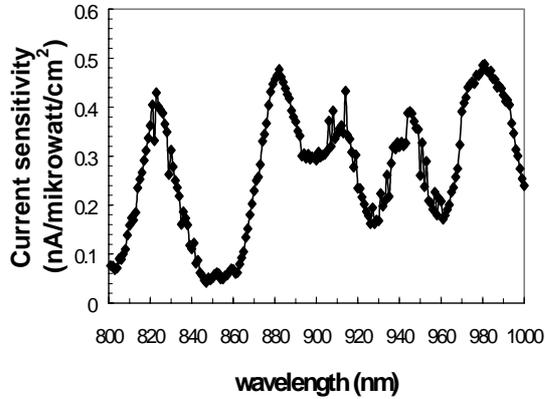

**Figure 4.** The current sensitivity versus IR wavelength at chopper frequency = 200 Hz.

The current sensitivity ($F_i$) is the ratio of output current ($I_p$) to the incident radiation power ($P_i$) as given by Eq. (4) : [6,7]

$$F_i = \frac{I_p}{P_i}. \qquad (4)$$

The incident radiation power at the surface was of several ranges from 20 $\mu$W/cm$^2$ to 30 $\mu$W/cm$^2$. The signals measured on the Al/Si photodiode were the power of the incident value measured signal, calculated quantum efficiency (using Eq. (3)) and current sensitivity value (using Eq. (4)) at chopper frequency are 200 Hz. The resulting curves are shown in Figure 3 and Figure 4. Regarding the spectral distribution depicted in this figure, it can see that sensitivity peaks appear when the measurement was performed using the ac mode. The peak was at 981 nm (IR wavelength), the incident radiation power ($P_i$) at the surface was 25 $\mu$W/cm$^2$, the current signal ($I_p$) was 12.2 nA, the quantum efficiency ($\eta$) was 6.16 % and the current sensitivity was 0.488 nA/$\mu$W/cm$^2$.

## 4. Conclusion

A new metal-semiconductor Al /Si photodiode the most interesting details consists in the spectral sensitivity over a broad range of IR wavelengths, from 800 nm to 1000 nm, if it is illuminated with continuos and discontinuous light. The sensitivity from IR of studied heterostructure makes it very attractive for IR sensor application. Also, this structure can be used for optoelectronic applications such as field effect controlled devices.

## ACKNOWLEDGMENT

This work was supported by Center Grant URGE Project, The Ministry of National Education, The Republic of Indonesia, under contract No. 008/CG//III/URGE/1997.